\documentclass[%
 reprint,
 amsmath,amssymb,
 aps,
]{revtex4-2}
\usepackage{lipsum}
\usepackage{graphicx}% Include figure files
\usepackage{dcolumn}% Align table columns on decimal point
\usepackage{bm}% bold math
\usepackage{color}
\usepackage{comment}
\usepackage{physics}
\usepackage{svg}
\usepackage{blindtext}
\usepackage{epsfig}
\begin{document}
\preprint{APS/123-QED}
\title{Self-duality properties and localization centers of the electronic wave functions  at high magic angles in twisted bilayer graphene}
\author{Leonardo A. Navarro-Labastida and Gerardo G. Naumis}
\date{September, 2023}
\email{naumis@fisica.unam.mx}
\affiliation{%
Depto. de Sistemas Complejos, Instituto de F\'isica, \\ Universidad Nacional Aut\'onoma de M\'exico (UNAM)\\
Apdo. Postal 20-364, 01000, CDMX, M\'exico.
}%
\begin{abstract}
Twisted bilayer graphene (TBG) is known for exhibiting highly correlated phases at magic angles due to the emergence of flat bands that enhance electron-electron interactions. The connection between magic angles and the Quantum Hall effect remains a topic of ongoing research. In the TBG chiral model, electronic wave function properties depend on a single parameter ($\alpha$), inversely proportional to the relative twist angle between the two graphene layers and which includes the interlayer interaction strength. In previous studies, as the twist angles approached small values, strong confinement  and a convergence to coherent Landau states were observed. However, the origin of these phenomena remained elusive. This work explores flat-band electronic modes, revealing that flat band states exhibit self-duality; they are coherent Landau states in reciprocal space and exhibit minimal dispersion, with standard deviation $\sigma_k=\sqrt{3\alpha/2\pi}$ as $\alpha$ approaches infinity. Subsequently, by symmetrizing the wave functions and considering the squared TBG Hamiltonian, the strong confinement observed in the $\alpha\rightarrow\infty$ limit is explained. This confinement arises from the combination of the symmetrized squared norm of the moiré potential and the quantized orbital motion of electrons, effectively creating a quantum well. The ground state of this well, located at defined spots, corresponds to Landau levels with energy determined by the magic angle. Furthermore, we demonstrate that the problem is physically analogous to an electron attached to a non-Abelian $SU(2)$ gauge field with an underlying $C_3$ symmetry. In regions of strong confinement, the system can be considered as Abelian, aligning with the picture of a simple harmonic oscillator.
This allows to define a magnetic energy in which the important role of the wave function parity and gap closing at non-magic angles is revealed. Finally, we investigate the transition from the original non-Abelian nature to an Abelian state by artificially changing the pseudo-magnetic vector components from an $SU(2)$ to a $U(1)$ field, which alters the sequence of magic angles.
\end{abstract}

\maketitle 
\section{Introduction}
Superconductivity in twisted bilayer graphene (TBG) is known to occur when the rotation angle between layers is able to produce a flat band in which electrons have zero group velocity \cite{Cao2018}.  Such angles are known as "magic angles." This important discovery has unveiled the significance of two-dimensional (2D) materials in understanding unconventional superconductivity in cuprates and heavy fermion systems, as they share similar quantum phase diagrams and present a new paradigm in moiré materials \cite{Cao2018, park2021, Bernevig2022}. After the discovery of superconductivity in TBG \cite{Cao2018}, other works reinforced the observation that flat bands are quite important to the existence of unconventional superconductivity and strongly correlated phases in twisted multilayer graphene systems \cite{2018ChengCheng, 2018Wu, Fidrysiak2018, 2019FengCheng, Yankowitz2019, You2019, Roy2019, Kerelsky2019, Uri2020, 2022Onari, LiangFu2021, CANO2021, KhalafEslam2021, Patrick2021, park2021, Ledwith2021, VafekOscar2021, PhongV2021, JohannesHofmann2021, Fernandes2021, Shen_2022, guerci2023_, patrickL2023}. \\
TBG flat bands, also known as zero mode states, share a lot of mathematical similarities to the ground state of the quantum Hall effect (QHE) \cite{Tarnpolsky2019, CanoJennifer2021, Naumis2023}. It was also known that magic angles exhibit a remarkable $3/2$ sequence or quantization rule, characterized by the vanishing of the Fermi velocity and the appearance of flat bands \cite{MacDonald2011,Tarnpolsky2019, CanoJennifer2021, Naumis2022,Naumis2023}.

G. Tarnoposky et. al. \cite{Tarnpolsky2019} found the simplest model for magic angles in TBG by turning off one of the hoppings between layers. This model was crucial for understanding the underlying symmetries such as intralayer inversion symmetry and the parity of magic angles. It also allowed for a deeper analysis of the structure of the zero mode wave function. \cite{CanoJennifer2021}. \\ 

Zero energy modes at magic angles have been investigated in many recent works \cite{2018ChengCheng, Tarnpolsky2019, Ledwidth2020, 2022Onari, Ledwith2021, WANGG2021, CANO2021, popovF2021, Saul2021, Patrick2021, 2023Tarnopolsky, 2023Grigory, Ledwidth2023, 2022Ledwith_Vortex}. There were mathematical hints for a possible connection with the QHE and the lowest Landau level \cite{Tarnpolsky2019, WANGG2021, Mera2021, Naumis2023, 2023MeraBruno}. Other works, revealed interesting connections with FQHE, topological matter, Weyl semimetals, Floquet systems, and anomalous edge states \cite{2018Low, 2018Wu, Xu2018, 2019FengCheng, Barth2020, Vogl2020, Pixley_2020, Stauber2020, 2021ShangL, 2021Liu, 2022Ledwith_Vortex, 2022Vishwanath, Nadia2022, PierrePantaleon2022}.

Working in magic-angle TBG it was indeed proved that the squared Hamiltonian of this system is closely related to the quantum harmonic oscillator and QHE \cite{Naumis2023}. The ground state is a flat band in which the wave function converges into coherent Landau-level states of the QHE. Another important result was the explanation of the mystery of the "3/2 magic angle recurrence rule" by using scaling arguments \cite{Naumis2023}. This rule is intimately related to the quantization of angular momentum. Consequently, for each magic angle, there exists a well-defined attached angular quantum number, which can be interpreted as interlayer currents \cite{Naumis2023}. 
This explanation of the basic principles underlying the magic-angle phenomenon provides valuable insights into addressing new fundamental questions at the intersection of the fractional quantum Hall effect (FQHE) and unconventional superconductivity. These questions are the subject of intense study in strongly correlated systems \cite{Jarrillo2021_Fractional}.

However, despite our previous works \cite{Naumis2021,Naumis2022,Naumis2023}, several questions remain unanswered. One of these questions pertains to the mechanism behind the strong localization of wavefunctions in magic-angle invariant spots once the lattice is properly scaled by the parameter $\alpha$, which encapsulates the energetic interaction coupling between layers and the angle. Additionally, we have yet to explore the consequences of nearly coherent Landau states. Here we show that zero modes behave as minimal dispersion packets as expected. We also explain how the wavefunction confinement arises around certain localization centers due to an effective potential produced by the moiré potential and the orbital motion of the electron. Moreover, we show that the
magic angle order parity is a crucial property associated with flat bands in twisted bilayer graphene. We also establish some connections between the angular momentum and non-Abelian pseudo-magnetic fields. \\

The present work is divided as follows. Section \ref{secCTBG} introduces the Hamiltonian for the chiral twisted bilayer graphene model and the pseudo-magnetic field that emerges due to the effect of the parameter $\alpha$. Section \ref{secSelfDuality} finds self-duality localization properties between reciprocal and real space and demonstrates that zero-mode states are coherent Landau states. Section \ref{secConfinement} analyzes confinement conditions for the electronic wavefunction in the asymptotic limit $\alpha\rightarrow\infty$ and the symmetries of the zero energy wavefunction. Section \ref{secNonAbelian} explores the non-Abelian nature of TBG and its connection with the magnetic QHE. Section \ref{secBeta} analyzes the non-Abelian nature of the pseudo-magnetic field by changing artificially its structure to make it more Abelian and how the scaling and recurrence are modified. Finally, section \ref{secConclusion} gives some conclusions and further research directions.

\section{Chiral Squared TBG Hamiltonian}\label{secCTBG}

The BM (Bistritzer-MacDonald) Hamiltonian was the first model to capture the nature of magic angle recurrence in TBG \cite{MacDonald2011}. Interestingly, taking $AA$ tunneling between layers equal to zero the spectrum in TBG has an extra chiral symmetry so, this reduced model is called the cTBG or TKV (Tarnopolsky-Kruchkov-Vishwanath) model. In the chiral basis, the bi-spinor is $\Phi(\bm{r})=\begin{pmatrix} 
\psi_1(\bm{r}) ,
\psi_2(\bm{r}),
\chi_1(\bm{r}),
\chi_2(\bm{r})
\end{pmatrix}^T$ where indexes $1,2$ denotes each graphene layer and $\psi_j(\bm{r})$ and $\chi_j(\bm{r})$ are the Wannier orbitals on each sub-lattice of the graphene's unit cell. 

The chiral Hamiltonian is given by \cite{Tarnpolsky2019,Khalaff2019, Ledwidth2020}, 
\begin{equation}
\begin{split}
\mathcal{H}
&=\begin{pmatrix} 
0 & D^{\ast}(-\bm{r})\\
 D(\bm{r}) & 0
  \end{pmatrix}  \\
\end{split} 
\label{H_initial}
\end{equation}
where the zero-mode operator is defined as, 
\begin{equation}
\begin{split}
D(\bm{r})&=\begin{pmatrix} 
-i\Bar{\partial} & \alpha U(\bm{r})\\
  \alpha U(-\bm{r}) & -i\Bar{\partial} 
  \end{pmatrix}  \\
\end{split} 
\end{equation}
with $\Bar{\partial}=\partial_x+i\partial_y$. The coupling potential between layers is,
\begin{equation}
    U(\bm{\bm{r}})=\sum^{3}_{\nu=1}e^{i\phi(\nu-1)}e^{-i\bm{q}_{\nu}\cdot \bm{r}}
\end{equation}
where the phase factor is $\phi=2\pi/3$ and the vectors are given by,

\begin{equation}
\begin{split}
    &\bm{q}_{1}=k_{\theta}(0,-1)\\
    &\bm{q}_{2}=k_{\theta}(\frac{\sqrt{3}}{2},\frac{1}{2})\\
    &\bm{q}_{3}=k_{\theta}(-\frac{\sqrt{3}}{2},\frac{1}{2})
\end{split}
\end{equation}
the moir\'e modulation vector is $k_{\theta}=2k_{D}\sin{\frac{\theta}{2}}$ with
$k_{D}=\frac{4\pi}{3a_{0}}$ is the magnitude of the Dirac wave vector and $a_{0}$ is the lattice constant of monolayer graphene. 
The cTBG model has only $\alpha$ as a parameter, defined as $\alpha=\frac{w_1}{v_0 k_\theta}$ where $w_1=110$ meV is the interlayer coupling of stacking AB/BA and $v_0=\frac{19.81eV}{2k_D}$ is the Fermi velocity. The diagonal operators $\partial$ and $\Bar{\partial}$ are dimensionless as eq. (\ref{H_initial}) is written in using units where $v_0=1$, $k_{\theta}=1$. The twist angle only enters in the dimensionless parameter $\alpha$ and scaling energy $\epsilon/\alpha$.

In $k$-space, the moir\'e Brillouin zone (mBZ) has 
\begin{equation}
\begin{split}
    &\bm{b}_{1,2}=\bm{q}_{2,3}-\bm{q}_{1}\\
    &\bm{b}_{3}=\bm{q}_{3}-\bm{q}_2
\end{split}
\end{equation}
as the moir\'e reciprocal vectors. Some important high symmetry points of the mBZ are $\bm{K}=(0,0)$, $\bm{K'}=-\bm{q}_1$, and $\bm{\Gamma}=\bm{q}_1$ \cite{Naumis2022}.  It is also convenient to define a set of unitary vectors $\bm{q}_{\nu}^{\perp}$ perpendicular to the set $\bm{q}_{\nu}$ and defined as,

\begin{equation}
\begin{split}
    &\bm{q}_{1}^{\perp}=(1,0)\\
    &\bm{q}_{2}^{\perp}=\big(-\frac{1}{2},\frac{\sqrt{3}}{2}\big)\\
    &\bm{q}_{3}^{\perp}=\big(-\frac{1}{2},-\frac{ \sqrt{3}}{2}\big)
\end{split}
\end{equation}
    
The moir\'e vectors  unitary cell are given by $\bm{a}_{1,2}=(4\pi/3k_{\theta})(\sqrt{3}/2,1/2)$. Note that $\bm{q}_{\nu}\cdot \bm{a}_{1,2}=-\phi$ for $\nu=1,2,3$.  In our previous works \cite{Naumis2021, Naumis2022, Naumis2023}, we demonstrated that squaring the Hamiltonian $\mathcal{H}$ allows us to simplify it into a $2 \times 2$ matrix that we call the squared Hamiltonian  $H^{2}$. In this work, we introduce notation changes in the definitions used inside $H^{2}$. The reasons will become evident later on. $H^{2}$ is given by,

\begin{equation}\label{eq:H2}
\begin{split}
&H^{2}=\\
&\begin{pmatrix} -\nabla^{2}+\alpha^{2}(\bm{A}^{2}+i[A_x, A_y] )&  \alpha (-2i\bm{A}_{-}\cdot\nabla +\nabla\times \bm{A}_{-}) \\
 \alpha (-2i\bm{A}_{+}\cdot\nabla +\nabla\times\bm{A}_{+}) & -\nabla^{2}+\alpha^{2}(\bm{A}^{2}-i[A_x, A_y]) 
  \end{pmatrix} 
  \end{split} 
  \end{equation}
where we defined, 
\begin{equation}
 \begin{split}
\bm{A}_{\pm}\equiv \bm{A}(\pm\bm{r})& =\sum_{\nu=1}^{3}e^{\pm i\bm{q}_{\nu}\cdot\bm{r}}\bm{q}_{\nu}^{\perp}\\
  \end{split} 
\end{equation}
here $\bm{A}_{\pm} $ is a pseudo-magnetic vector potential with $C_3$ symmetry and $\bm{A}^{2}=|\bm{A}_{\pm}|^{2}$.  The squared norm of the coupling potential is an effective intralayer confinement potential,
\begin{equation}
\begin{split}
|U(\pm\bm{r})|^{2} &= \bm{A}^{2}\mp i[A_x, A_y]
\end{split}
\end{equation}
where the confinement potential $|U(\pm\bm{r})|^{2}$ is separated into its purely symmetric $\bm{A}^2(\bm{r})$ and anti-symmetric $i[A_x, A_y]$ parts defined as, 

\begin{equation}
\begin{split}
\bm{A}^2(\bm{r})= 3-\sum_{\nu}\cos{(\bm{b}_{\nu}\cdot\bm{r})}\\
\Delta(\bm{r})=\sqrt{3}\sum_{\nu}(-1)^{\nu}\sin{(\bm{b}_{\nu}\cdot\bm{r})}
\label{eq:potentials_s_anti}
\end{split}
\end{equation}
here $\Delta(\bm{r})=i[A_x,A_y]$ where $A_x$ and $A_y$ are the non-Abelian components of the $SU(2)$ pseudo-magnetic vector potential (See Appendix A). It is important to remark that the pseudo-magnetic vector potential satisfies the relation $\gradient\cdot\bm{A}_{\pm}=0$, so is a Coulomb gauge invariant field and $\gradient\times\bm{A}_{+}=\bm{B}_{+}$ (layer 1) and $\gradient\times\bm{A}_{-}=\bm{B}_{-}$  (layer 2).
The magnetic field is thus given by,
\begin{equation}
    \bm{B}(\pm\bm{r})=\pm i \sum_{\nu}e^{\pm i\bm{q}_{\nu}\cdot\bm{r}}\bm{e}_z
\end{equation}
where we have used the identity $\bm{e}_z=\bm{q}_{\nu}\times\bm{q}^{\perp}_{\nu}$ and $\bm{e}_z$ is a unitary vector in the direction perpendicular to the graphene's plane.

Notice that squaring the chiral TBG model is akin to a supersymmetric transformation \cite{DikiMatsumoto2023, Hatsugai2020, TomonariMizoguchi2022, Yoshida2021,2023Mizoguchi},  which seems to play a role in the proposed equivalence between the squared TBG electron Hamiltonian and an electron coupling to a $SU(2)$ non-Abelian pseudo-magnetic field \cite{Naumis2021}.

\section{Self-duality properties and convergence into coherent Landau states}\label{secSelfDuality}
%Recent progress demonstrated that the twisted bilayer graphene wave function has a similar structure to the quantum Hall effect wave function, indeed, this discovery was significant in finding interesting properties and analogies like fractional quantum hall states, topological phases, and Landau levels.  

It has been demonstrated that twisted bilayer graphene has Landau levels \cite{CANO2021, JieWang2023, Ledwidth2023, patrickL2023}. They play a crucial role in its remarkable properties like superconductivity, fractional Chern insulator phases \cite{KhalafEslam2021, Jarrillo2021_Fractional, 2021yardenn, 2021ShangL, 2022Vishwanath, Ledwith2021, 2022Ledwith_Vortex, 2022Ashvin}. However, there are some gaps related to the understanding of electronic localization in TBG from the perspective of one particle. For example, why for $\alpha\rightarrow\infty$ does the wavefunction localize at specific regions in real space and $k$-space? and how both spaces relate?

In a recent previous paper we demonstrated that the wave function in TBG exhibits an almost coherent Landau state nature with a dispersion $\sigma=1/\sqrt{3\alpha}$ which is only reached in the asymptotic limit \cite{Naumis2023}. This asymptotic limit squeezes the bands and makes these theoretically coherent states difficult to measure but here we are not worried about such a fact at this moment. We are more concerned about making some analogies and connections with Landau levels. Here we are going to discuss some properties of the wave functions and their relationship with coherent states. 

Coherent states are self-dual in the sense that their Fourier transforms in reciprocal space look similar to the real space but with inverted parameters. As a consequence, they satisfy the minimal uncertainty relation between real and momentum space. Let us now explore if such property is valid for TBG zero modes.

\begin{figure}[h!]
\includegraphics[scale=.45]{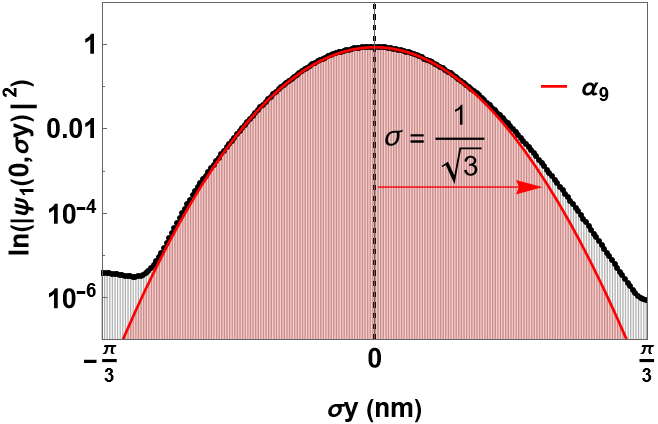}\caption{Electronic density, in log scale, of the higher magic angle $\alpha_9$ for the $\bm{\Gamma}$-point, and as a function of the position along $y$-axis. Black points are the numerical data obtained from the Hamiltonian. A normalized $y'$ variable was used such that $y^{\prime}=(y-1.047)/\sqrt{\alpha}$. The red curve is a Gaussian fit for $\psi_{1}(\bm{r})$. Notice the fat tails of the electronic density when compared with a Gaussian.}
\label{fig:gaussian_real}
\end{figure}

As seen in Fig. \ref{fig:gaussian_real}, the electronic probability density in real space for the ninth magic angle $\alpha_9$, with normalized coordinates as $\frac{y-\bm{R}}{\sqrt{\alpha}}$, where $\bm{R}\approx 1.047 $ is the position of one of the numerically found maximums (this value suggests that $\bm{R}\approx \pi/3$ but we do not have a proof of this conjecture), is almost a Gaussian. For comparison, in Fig. \ref{fig:gaussian_real} we plot a Gaussian with the same dispersion. Fig. \ref{fig:gaussian_real} reveals that the electronic distribution has a power-law fat tail decay. Interestingly, this makes the electronic density somewhat similar to the velocity-distribution fluctuations in turbulence \cite{CASTAING1990177}.

\begin{figure}[h!]
\includegraphics[scale=.4]{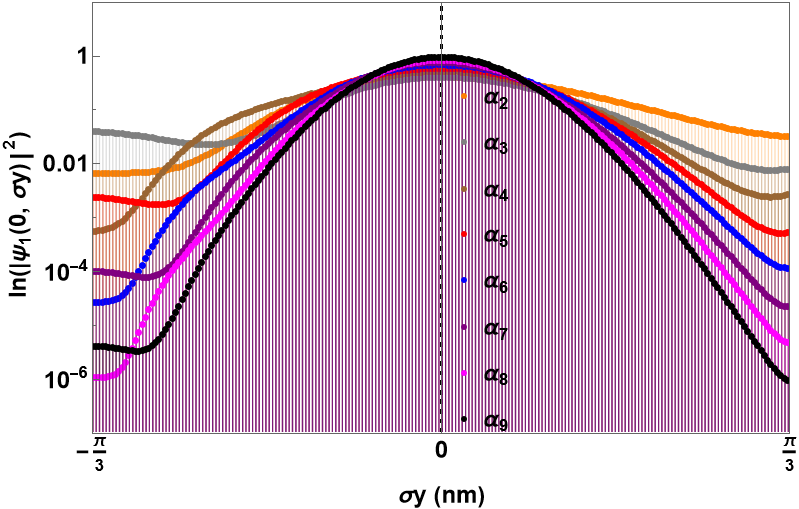}\caption{Electronic density, in log scale, from the second to ninth magic angles for the $\bm{\Gamma}$-point and as a function of the position along the $y$-axis. For simplicity, a normalized $y'$ variable was used such that $y^{\prime}=(y-1.047)/\sqrt{\alpha}$. Notice the convergence into a Gaussian.}
\label{fig:gaussian_real_magics}
\end{figure}

However, $\alpha$ squeezes these fat tails as this scaling parameter increases. This is shown in Fig. \ref{fig:gaussian_real_magics} where we plot the electronic probability in real space from the second to the ninth magic angles written in normalized coordinates, i.e., with zero mean and standard deviation one. Clearly, as the system goes to higher magic angles the fat tail diminishes and asymptotically converges to an invariant Gaussian distribution. 

%[h!]

As the positions of maximal electronic density probability near the origin are located  at $\bm{R} \approx \pm 1.047 \bm{q}_{\nu}$, the density can be approximated by a Gaussian distribution near $\bm{R}$ as,
\begin{equation} 
\begin{split}  
|\psi(\bm{r})|^2\approx \frac{3 A_M}{2\pi\sigma}e^{-\frac{1}{2\sigma^2}|\bm{r}\pm\bm{R}|^2}
\end{split}  
\label{eq:Gaussian_r} 
\end{equation} 
where $A_M=8\pi^2/(3\sqrt{3})$ is the normalized moiré unit cell area and $\sigma=1/\sqrt{3\alpha}$ is the standard deviation. Note that eq. (\ref{eq:Gaussian_r}) is independent of $\alpha$. To include the fat tails, we can use another function $W_{\alpha}(\bm{r})$ which is $\alpha$ dependent such that,
\begin{equation} 
\begin{split}  
|\psi(\bm{r})|^2\approx \frac{ A_M}{2\pi\sigma}e^{-\frac{1}{2\sigma^2}|\bm{r}\pm\bm{R}|^2}|W_{\alpha}(\bm{r})|^{2}
\end{split}  
\label{eq:Gaussian_r1} 
\end{equation} 
in agreement with other works \cite{JieWang2023,Popov2020}.
These fat tails are interesting as they allow to produce wave function overlaps though, at the same time,  are strongly localized in certain regions. \\ 

Coherent states have the property of being minimal dispersion wave packets. We explore this property for TBG by looking at the reciprocal space. As the wave functions follow Bloch's theorem, they can be written as \cite{Tarnpolsky2019}, 

\begin{equation}
\begin{split}\label{eq:spinor2}
\Psi_{\bm{k}}(\bm{r})=\begin{pmatrix} \psi_{\bm{k},1}(\bm{r})\\
\psi_{\bm{k},2}(\bm{r})
  \end{pmatrix} =\sum_{l,n}\begin{pmatrix} a_{ln}\\
 b_{ln}e^{i\bm{q}_1\cdot\bm{r}}
  \end{pmatrix}e^{i(\bm{K}_{ln}+\bm{k})\cdot\bm{r}}
  \end{split} 
  \end{equation}
where $a_{ln}$ and $b_{ln}$ are Fourier coefficients for layer 1 and layer 2 respectively.  $\bm{k}$ is a generic reciprocal wave vector and $\bm{K}_{ln}=l\bm{b}_1+n\bm{b}_2$. The vectors $\bm{b}_{1}=(\frac{\sqrt{3}}{2},\frac{3}{2})$ and  $\bm{b}_{2}=(-\frac{\sqrt{3}}{2},\frac{3}{2})$ are the two Moiré Brillouin zone vectors defined in section \ref{secCTBG}. 

\begin{figure}[h!]
\includegraphics[scale=.38]{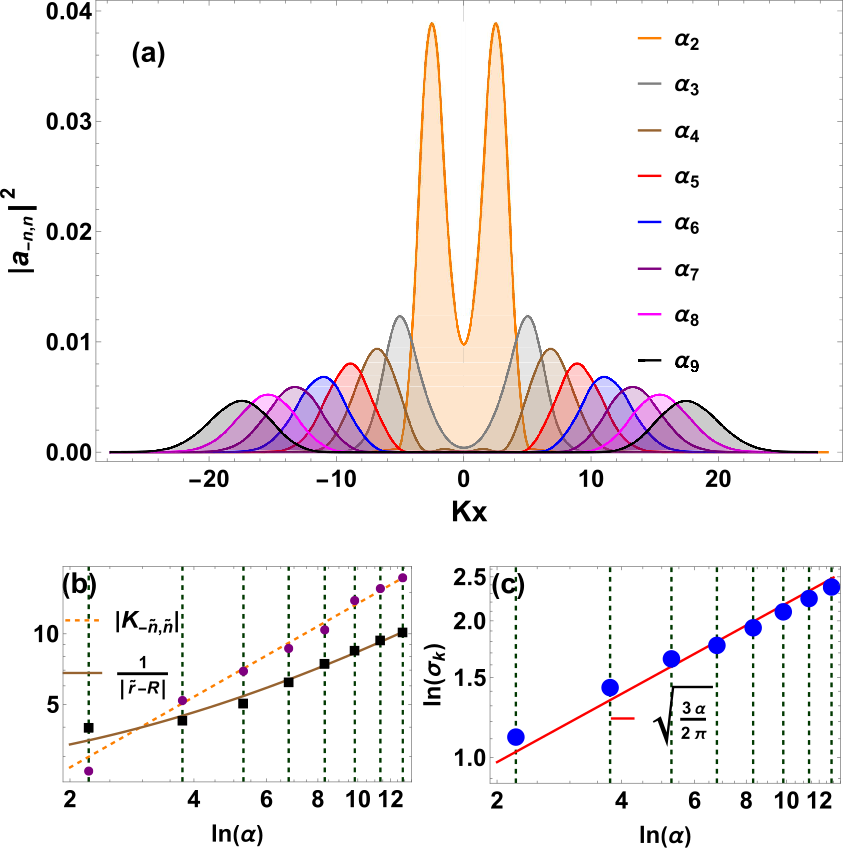}\caption{Fourier coefficients in reciprocal space in the direction $\bm{K}_{-\Tilde{n},\Tilde{n}}=\Tilde{n}\bm{b}_{3}$. Panel (a) shows the squared norm of Fourier coefficients $|a_{-n,n}|^2$ from the second to the ninth magic angles along the direction $\bm{K}_x=n(\bm{b}_2-\bm{b}_1)$. Panel (b) presents the convergence, in log-log scale, for the values $|K_{-\Tilde{n},\Tilde{n}}|$ (purple dots) and $1/|\Tilde{\bm{r}}-\bm{R}|$ (black squares) with $\bm{R}\approx 1.047 \bm{q}_1$. The associated lines for each marker are the linear fits $|K_{-n,n}|\approx 1.34\alpha$ (orange dashed) and $1/|\Tilde{\bm{r}}-\bm{R}|\approx 2.12271+0.626839\alpha$ (brown solid). Panel (c) shows the standard deviation in the log-log scale for the Gaussian distribution at the maximum point $\bm{K}_{-\Tilde{n},\Tilde{n}}$. Here it is numerically proved that $\sigma_k=\sqrt{\frac{3\alpha}{2\pi}}$ in $k$-space with the relation $\sigma_k=1/(\sqrt{2\pi}\sigma_r)$, where the indexes $k$ and $r$ represents $k$-space or real-space, respectively. This result shows that solutions are coherent states because they minimize the dispersion $\sigma_r\sigma_k=1/\sqrt{2\pi}$ thus, with minimal uncertainty relation $\sigma^2_r\sigma^2_k=\hbar$, where $\hbar=h/2\pi$ using natural units $h=1$ as the Plank's constant. }
\label{fig:gaussian_k}
\end{figure}

In Fig. \ref{fig:gaussian_k} panel (a) we present the Fourier coefficients squared norm for the zero mode wave function at the $\Gamma$ point for $\bm{K}_{x}=n(\bm{b}_2-\bm{b}_1)$, given by $|a_{-n,n}|^{2}$, for magic angles between $\alpha_2$ to $\alpha_9$. We can clearly see the Gaussian shape of the peaks, which turn out to be similar to the wave function in real space seen in Fig. 2 of our previous work \cite{Naumis2023}. This is in agreement with the idea of states converging into coherent states. As we can see, the coefficients $|a_{-n,n}|^{2}$ for $\alpha_2$ are strongly localized while for higher magic angles $\alpha_9$, the two original mirrors symmetric Gaussian's are quite separated, while the dispersion increases. For the real space case, the situation is reversed because the Gaussian's are more localized and their dispersion is reduced for higher magic angles (See Ref. \cite{Naumis2023}). In Fig. \ref{fig:gaussian_k} panel (b), we show the peak position of the Gaussian in $k$-space ($|\bm{K}_{-\Tilde{n},\Tilde{n}}|$), were $(-\Tilde{n},\Tilde{n})$ correspond to the reciprocal point with maximal norm Fourier coefficient, i.e., the positions of the maximums in reciprocal space along one direction. This is compared with the inverse of the difference between the wave function peaks positions in real space ($\tilde{\bm{r}}$) and the limiting localization center for $\alpha \rightarrow \infty$, i.e., we plot $1/|\tilde{\bm{r}}-\bm{R}|$. \\  

On the other hand, panel (c) presents the dispersion in $k$-space, denoted by $\sigma_{k}$, as a function of $\alpha$, showing that the dispersion increases with $\alpha$. This is easy to explain. Considering that $\psi(\bm{r})$ are almost coherent states, in a previous work \cite{Naumis2023} we showed that the dispersion in real space converges to $\sigma=1/\sqrt{3\alpha}$. Therefore, using that the Fourier transform of a Gaussian is another Gaussian with inverse standard deviation, we obtain that the dispersion in reciprocal space goes as,
\begin{equation}
    \sigma_k=\sqrt{\frac{3\alpha}{2\pi}}
\end{equation}
in agreement with  Fig. \ref{fig:gaussian_k} panel (c). Both in  Fig. \ref{fig:gaussian_k} panels (b) and (c), the vertical lines indicate magic angles. The solid lines are the theoretical results and the markers are the numerical results. We use the log-log scale for visual convenience. From these results, we can conclude that indeed our states converge into coherent states because they satisfy Heisenberg's uncertainty relation with minimal dispersion, i.e., 

\begin{equation} 
\begin{split}  
\sigma_r\sigma_k \approx \sqrt{\frac{1}{3\alpha}}\sqrt{\frac{3\alpha}{2\pi}}  =\sqrt{\frac{1}{2\pi}} 
\end{split}  
\label{eq:heisenberg} 
\end{equation} 
\\
or using natural units $h=1$ (Plank's constant) we end with, 
\begin{equation} 
\begin{split}  
\Delta_r\Delta_k \approx \hbar 
\end{split}  
\label{eq:heisenberg2} 
\end{equation} 
where $\Delta_r=\sigma^2_r$ and $\Delta_k=\sigma^2_k$. The result $\hbar$ is a consequence of the model because we are treated with a $2D$ model and each degree of freedom contributes $\hbar/2$ to the dispersion, in analogy to a $2D$ quantum harmonic oscillator.\\

\begin{figure}[h!]
\includegraphics[scale=.235]{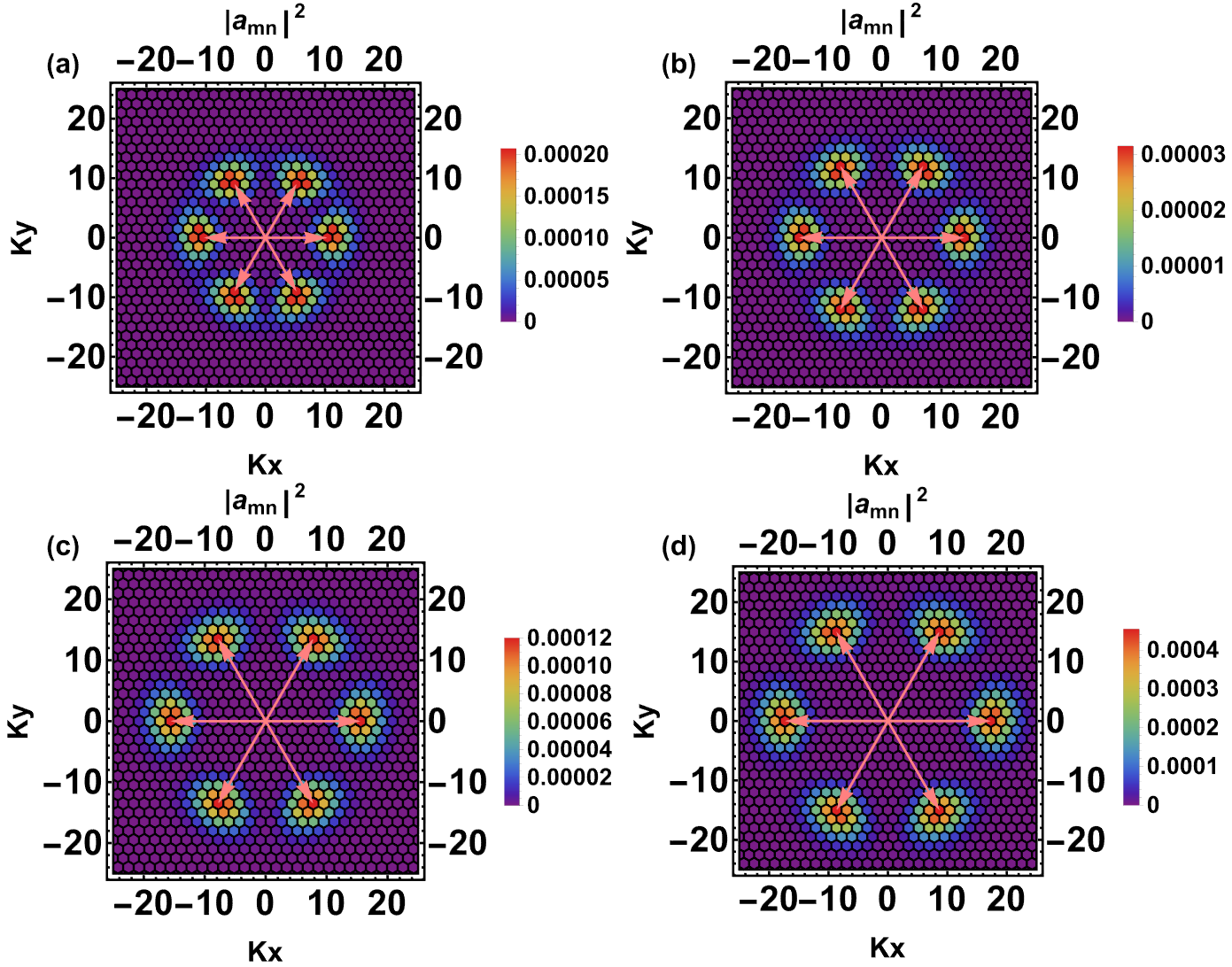}\caption{Fourier coefficients squared norm color map for the zero-mode wavefunction for high magic angles. Panel (a), $\alpha_6=8.313$, (b) $\alpha_7=9.829$, (c) $\alpha_8=11.345$ and (d) $\alpha_9=12.855$. All correspond to the $\Gamma$-point coefficients. The arrows indicate the positions of the maximal norm Fourier coefficients, and are the centers of the coherent Landau states in reciprocal space. The centers are located at $\tilde{n}\bm{b}_{\nu} \pm 1.047\bm{q}_{\nu}$ where $\tilde{n}\approx\sqrt{3}\alpha_{m}/2$ for $m\rightarrow\infty$ higher magic angles, and $ C_3$ rotations relate produce the extra points seen in the figure. Observe how as the magic angle order grows, the maxima are pushed away from the center.}
\label{fig:coefficients}
\end{figure}

To give more insight into the localization centers in reciprocal space,  Fig. \ref{fig:coefficients} presents a color map for the Fourier coefficients $|a_{mn}|^2$ (layer 1) for the $\Gamma$-point wave function. From panel (a) to panel (d) the magic angle order increases and the maxima of the Fourier coefficients departs radially from the center. Pink arrows indicate where the sixth localization center lies. 

According to these numerical results,  the maximums of the electronic probability in k-space are near,
\begin{equation} 
\begin{split}  
\Tilde{n}\bm{b}_{\nu}\pm 1.047\bm{q}_{\nu}
\end{split}  
\label{eq:vectors} 
\end{equation} 
and their corresponding rotated versions by $2\pi/3$. In real space, the maxima are at,
\begin{equation}
    \bm{R} \approx \frac{1}{\Tilde{n}}\hat{\bm{R}}_{-\phi}(\bm{b}_{\nu})+1.047\hat{\bm{R}}_{-\phi}(\bm{q}_{\nu})
\end{equation}
Here $\hat{\bm{R}}_{-\phi}$ represents a rotation by an angle $\phi=\frac{2\pi}{3}$ and $\tilde{n}\approx\sqrt{3}\alpha_{m}/2$. For the other layer, the same behavior occurs with the Fourier coefficients ($|b_{mn}|^2$). Therefore, we can summarize such behavior as follows. As $\alpha \rightarrow \infty$, wave functions become strongly confined in certain spots. In reciprocal space, the confinement is also present but decreases with growing $\alpha$ and at the same time, the location of the maximums goes to infinity. To delve deeper into such properties, in the following section we discuss how and why confinement at certain locations arises.

\section{Confinement and wave function symmetries}\label{secConfinement}

As was discussed in the previous section and in previous works \cite{Naumis2022,Naumis2023}, the wave functions in real space converge into very sharp Gaussian packets which are located at
the invariant points $\bm{R}$. In this section, we discuss the origin of this effect as well as some symmetry properties of the wave function required to understand how the confinement arises.  
Let us show first how at higher magic angles the wave function in real space can be decoupled into symmetric and anti-symmetric parts. These are spatially located at different regions and depend on the magic angle order parity. To clarify these points, it is convenient to write the zero-mode equation of the squared Hamiltonian, 
\begin{equation}
\begin{split}
(-\laplacian+\alpha^2(&\bm{A}^{2}+i[A_x, A_y]))\psi_{1}(\bm{r})\\
&+\alpha(-2i\bm{A}_{-}\cdot\nabla +\nabla\times \bm{A}_{-})\psi_{2}(\bm{r})=0
    \label{eq:zero_mode_h2}
\end{split}
\end{equation}
At this point we remark that the eigenfunctions of $\mathcal{H}$ are simultaneously eigenfunctions of $H^2$, however, the vise-verse is not. Here we will work with $H^{2}$ because has more physical relevance for the present discussion, however, the numerical calculations of the wave function that we will present in what follows are in the $4\times4$ chiral basis of $\mathcal{H}$. As explained elsewhere \cite{Naumis2022}, any linear combination of degenerate eigenfunctions of  $\mathcal{H}$ are solutions of $H^{2}$, so there is a phase involved. In spite of this, the electronic density and energy contributions are not affected if they are calculated in $H^2$ or $\mathcal{H}$ as the phase factor is eliminated. \\

For simplicity, in this analysis, we will first consider the $\Gamma$-point. In this case the symmetry allows to write $\psi_{2}(\bm{r})=i\mu_{\alpha}\psi_{1}(-\bm{r})$ with $\mu_{\alpha}=\pm 1$ as the magic angle order parity \cite{Tarnpolsky2019}. For odd parity magic angle order, i.e, for $\alpha_{2m+1}$ we have $\mu_{\alpha}=+1$, while for even parity ($\alpha_{2m}$) $\mu_{\alpha}=-1$. 

\begin{figure}[htp]
\includegraphics[scale=.29]{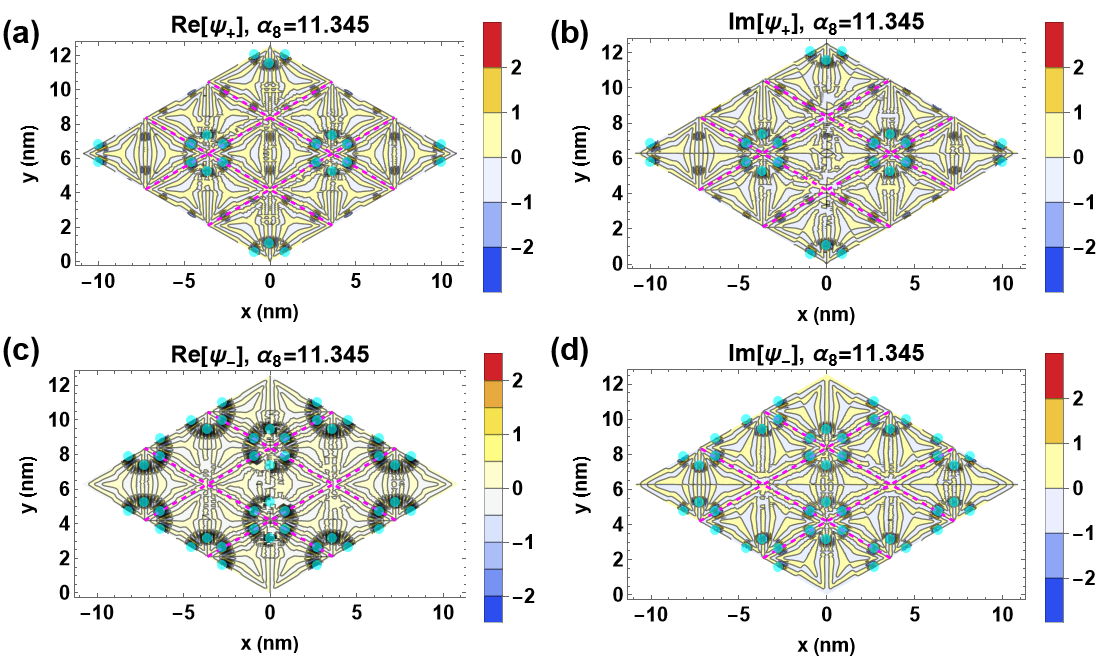}\caption{Symmetric ($\psi_{+}(\bm{r})$) and anti-symmetric($\psi_{-}(\bm{r})$) wave functions in a $3\times 3$ unit cell for $\alpha_8=11.345$. The blue circles indicate where the electronic wave function is localized and the dashed lines show unit cells defined by the vectors $\bm{a}_1$ $\bm{a}_2$. Symmetric/anti-symmetric wave functions are defined as $\psi_{\pm}=\psi_{1}(\bm{r})\mp i\mu_{\alpha}\psi_{2}(\bm{r})$. Considering the $\Gamma$-point $\psi_{2}(\bm{r})=i\mu_{\alpha}\psi_{1}(-\bm{r})$ symmetric/anti-symmetric solutions changes as $\psi_{\pm}=\psi_{1}(\bm{r})\pm\psi_{1}(-\bm{r})$. (a-b) Real and imaginary parts of the symmetric wave function $\psi_{+}$. (c-d) Real and imaginary parts of the anti-symmetric wave function $\psi_{-}$. Note that symmetric and anti-symmetric solutions are almost spatially decoupled.}
\label{fig: celda1}
\end{figure}

\begin{figure}[htp]
\includegraphics[scale=.2755]{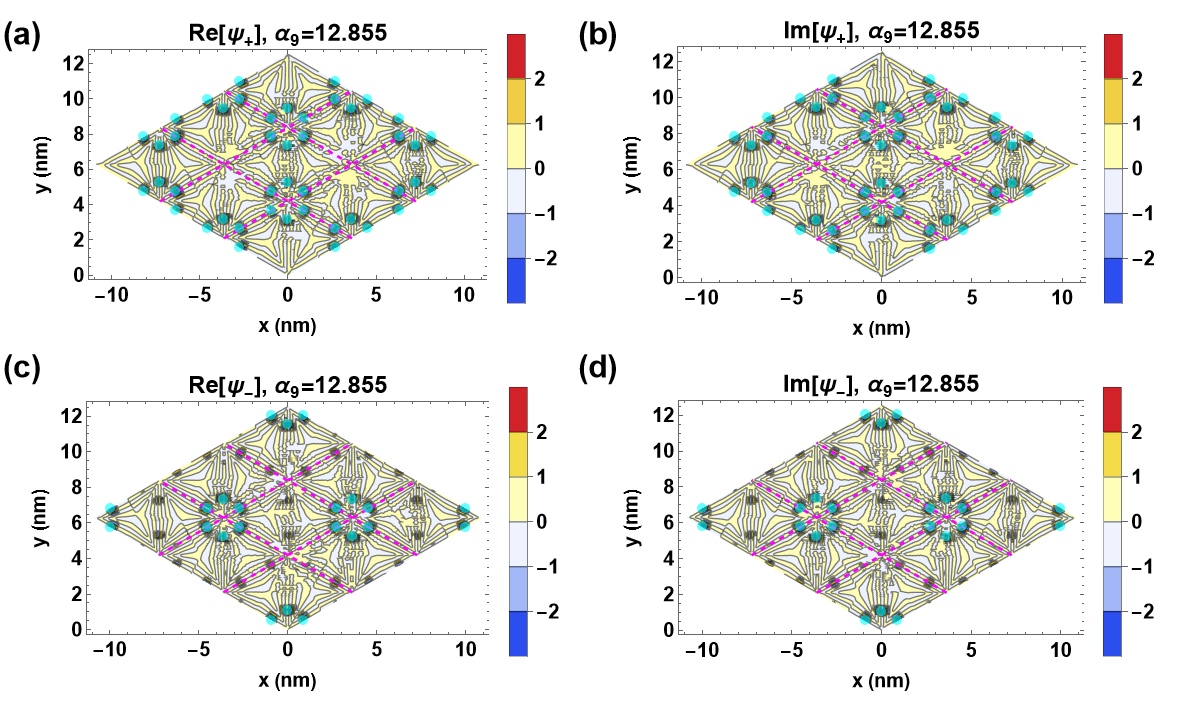}\caption{Symmetric and anti-symmetric wave functions in a $3\times 3$ unit cell for $\alpha_9=12.855$. The blue circles indicate where the electronic wave function is localized and the dashed lines show unit cells defined by the vectors $\bm{a}_1$ $\bm{a}_2$. Symmetric/anti-symmetric wave functions are defined as $\psi_{\pm}=\psi_{1}(\bm{r})\mp i\mu_{\alpha}\psi_2(\bm{r})$. Considering the $\Gamma$-point $\psi_2(\bm{r})=i\mu_{\alpha}\psi_1(-\bm{r})$ symmetric/anti-symmetric solutions changes as $\psi_{\pm}=\psi_{1}(\bm{r})\pm\psi_1(-\bm{r})$. (a-b) Real and imaginary parts of the symmetric wave function $\psi_{+}$. (c-d) Real and imaginary parts of the anti-symmetric wave function $\psi_{-}$. Note that symmetric and anti-symmetric solutions are almost decoupled.}
\label{fig: celda2}
\end{figure}

We now define symmetric or anti-symmetric functions as $\psi_{\pm}(\bm{r})=\psi_{1}(\bm{r})\pm\psi_{1}(-\bm{r})$. Therefore, the pair of zero mode eqn. (\ref{eq:zero_mode_h2}) can be rewritten as, 

\begin{equation}
\begin{split}
(-\laplacian&+\alpha^{2}\bm{A}^2-i\mu_{\alpha}\alpha( -2i\mathcal{A}_{\mp}\cdot\nabla+\nabla\times\mathcal{A}_{\mp}))\psi_{\pm}\\
&+(\alpha^{2}\Delta-i\mu_{\alpha}\alpha ( -2i\mathcal{A}_{\pm}\cdot\nabla+\nabla\times\mathcal{A}_{\pm}))\psi_{\mp}=0
\label{eq:zero_mode_h2_symmetric_2}
\end{split}
\end{equation}
where we also defined the symmetry/anti-symmetry non-Abelian pseudo-magnetic field as,
\begin{equation}
\mathcal{A}_{\pm}=(\bm{A}_{+}\pm \bm{A}_{-})/2
\label{eq:A_symmetric_operator}
\end{equation}

Our numerical results in Fig. \ref{fig: celda1} and Fig. \ref{fig: celda2} highlight that indeed the solutions are decoupled spatially in this symmetric or anti-symmetric basis. For example, in Fig. \ref{fig: celda1} the magic angle ($\alpha_8=11.345$) has even order parity ($m=8$) with $\mu_{\alpha}=-1$. In panels (a)-(b) we present the real and imaginary parts respectively of the symmetric solution $\psi_{+}$. The blue dots indicate the corresponding maxima.  In panels (c)-(d) we present a similar plot for $\psi_{-}$. The maxima of $\psi_{-}$ are in different locations than those in $\psi_{+}$. Moreover, for even parity, the anti-symmetric solution doubles the number of maxima when compared with the symmetric solution. Quite remarkably, if we continue with the next magic angle, the parity changes to an odd magic angle ($\alpha_9=12.855$) with $\mu_{\alpha}=+1$. Note that in Fig. \ref{fig: celda2} the situation is reversed, now  $\psi_{+}$ has the double of peaks when compared with  $\psi_{+}$. The localization centers of $\psi_{+}$ and $\psi_{-}$ are interchanged when compared with $\alpha_8$. \\

Observe how both in Fig. \ref{fig: celda1}-\ref{fig: celda2}, magenta dashed lines indicate moiré unit cells while the supercell here is $3\times3$ bigger as the pseudo-magnetic potentials define a bigger magnetic unit cell \cite{Naumis2023}. This bigger period is seen in the coupling potential as $U(\bm{r}+\bm{a}_{1,2})=e^{-i\phi}U(\bm{r})$, thus this requires a translation of $3\bm{a}_{1,2}$ to recover the crystal periodicity and a phase factor $e^{3i\phi}=1$. In such a bigger unit cell, the potential is periodic and in fact, leads to the quantization rule for the magic angles \cite{Naumis2023}. The  $3\times3$ unitary cells are essential to clearly understand the inversion symmetries of the wave functions as if only one unitary moiré cell is used, defined by $\bm{a}_{1,2}$, the extra phases make the interpretation very difficult.\\
\\
Our numerical results indicate distinct localization regions for $\psi_{+}$ and $\psi_{-}$, suggesting that in equation (\ref{eq:zero_mode_h2_symmetric_2}), each term can be separately set to zero to satisfy the equation, owing to the strong confinement. Thus, as a solution, we propose that eq. (\ref{eq:zero_mode_h2_symmetric_2}) can be decoupled into, 

\begin{equation}
(-\laplacian+\alpha^{2}\bm{A}^2-i\mu_{\alpha}\alpha( -2i\mathcal{A}_{\mp}\cdot\nabla+\nabla\times\mathcal{A}_{\mp}))\psi_{\pm} \approx 0
\label{eq:zero_mode_h2_symmetric_decoupled}
\end{equation}
\begin{equation}
(\alpha^{2}\Delta-i\mu_{\alpha}\alpha( -2i\mathcal{A}_{\pm}\cdot\nabla+\nabla\times\mathcal{A}_{\pm}))\psi_{\mp} \approx 0
\label{eq:zero_mode_h2_symmetric_decoupled2}
\end{equation}

%and a scaling of the spatial coordinates as,
%\begin{equation}
%\begin{split}
%\bm{r}^{\prime}=\bm{r}/\alpha
%\label{eq:scaling_argument}
%\end{split}
%\end{equation}

%then $\gradient^{\prime}=(\alpha\gradient)$ and $(\gradient^{\prime})^{2}=(\alpha\gradient)^2$, 

As explained in Appendix B, by using eqns. (\ref{eq:zero_mode_h2_symmetric_decoupled}) and (\ref{eq:zero_mode_h2_symmetric_decoupled2}) it can be proved that the following eq. is obtained,
\begin{equation}
(-\laplacian+\alpha^{2}\bm{A}^2(\bm{r})-\alpha^{2}\Delta(\bm{r}))\psi_{\pm}\approx 0
\label{eq:zero_mode_h2_symmetric_poisson}
\end{equation}
where in eq. (\ref{eq:zero_mode_h2_symmetric_poisson}) it is supposed $\alpha\rightarrow\infty$ and thus $\nabla\times\mathcal{A}_{\pm}(\bm{r})\rightarrow 0$ is negligible as it scales as $\alpha$.  This indeed supports the  use of well-defined parity wave functions as was done in a previous work \cite{Naumis2023}. 

\begin{figure}[h!]
\includegraphics[scale=.45]{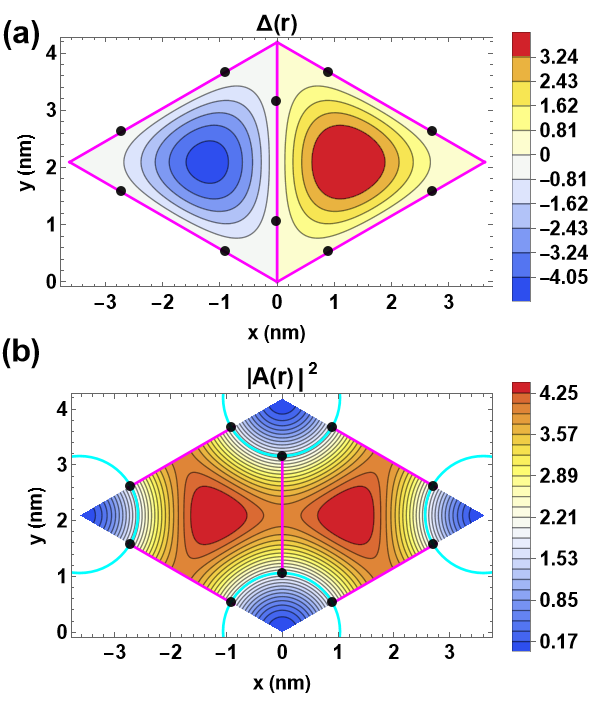}\caption{Confinement spots and potentials in the unit cell defined using the vectors $\bm{a}_1$ and $\bm{a}_2$. (a) Anti-symmetric potential $\Delta(\bm{r})$ and (b) symmetric potential $\bm{A}^2(\bm{r})$. The black points are the localization centers of the electronic zero-mode wave function. In the anti-symmetric potential $\Delta(\bm{r})$, magenta lines indicate angular confinement directions where locally the non-Abelian commutator is zero, $\Delta(\pm 1.047 \bm{q}_{\nu})= i[A_{x}, A_{y}]=0$ the directions are defined by vectors $\pm 1.047\bm{q}_{\nu}$. The symmetric potential $\bm{A}^2(\bm{r})$ is also important because it tells us some information related to radial confinement. In (b), cyan circles have a radius $1.047$, and black points lie around these circles. More importantly, $\bm{R} \approx \pm 1.047 \bm{q}_{\nu}$ corresponds to special points restricted by the angular confinement directions of $\Delta(\bm{r})$. These special points are also related to tunneling paths (magenta lines) that are energetically favorable and connect electronic density centers by a saddle point.}
\label{fig:potentialA}
\end{figure}

As is seen in eq. (\ref{eq:zero_mode_h2_symmetric_poisson}), the potential $\bm{A}^2(\bm{r})-\Delta(\bm{r})$ governs the electronic localization behavior in the asymptotic limit $\alpha\rightarrow\infty$. However, note that taking $\bm{r}\rightarrow -\bm{r}$ in eq. (\ref{eq:zero_mode_h2_symmetric_poisson}) changes the sign of $\Delta(-\bm{r})=-\Delta(\bm{r})$ while keeping invariant the other terms. This property allows for the decoupling of the symmetric and anti-symmetric potentials as, 
\begin{equation}
\begin{split}
(-\laplacian+\alpha^{2}\bm{A}^2(\bm{r}))\psi_{\pm}\approx 0\\
\Delta (\bm{r}) \psi_{\pm}\approx 0
\label{eq:zero_mode_h2_symmetric_poisson2}
\end{split}
\end{equation}

\begin{figure}[h!]
\includegraphics[scale=.5]{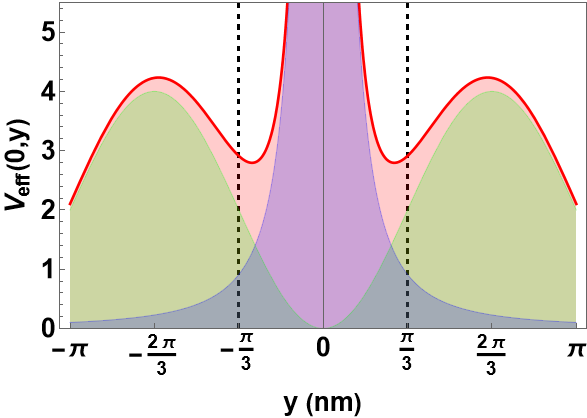}\caption{Effective potential $V_{eff}(r)$ along the axis $\bm{r}=(0,y)$. The blue curve is the function $1/y^2$ while the green curve is $A^2(0,y)$. Electrons are confined in the well around the local minima of the effective potential at $R \approx 0.88 $. In this plot, we include two dashed vertical lines that indicate the position where the numerically found electronic wave function has its localization center ($R \approx 1.047$) for the limit $\alpha\rightarrow\infty$. }
\label{fig:potencial_eff}
\end{figure}

To satisfy the second of the previous equations, we must have $\Delta(\bm{r}) \approx 0$ in regions where  $\psi_{\pm} \ne 0$. Fig. \ref{fig:potentialA} (a) confirms numerically that such condition is correct, i.e., wave functions are localized in the lines for which  $\Delta(\bm{r})=0$. Moreover, this implies that localization occurs whenever $[A_x, A_y]=0$. Therefore, locally the system is Abelian. 
 As shown in  Appendix A, the positions where $\Delta(\bm{r})=0$ occur at high-symmetry directions so the localization centers, for the vertex at the origin,  will have numerically found positions near,
 \begin{equation}
     \bm{R}\approx \pm R\bm{q}_{\nu}
     \label{eq:guiding_centers}
 \end{equation}
where $R=1.047...$ is the magnitude of $\bm{R}$. It gives the radial distance of the maximum to the vertex of the cell. Its value is determined from the condition $(-\laplacian+\alpha^2\bm{A}^2(\bm{r}))\psi_{\pm}\approx 0$. 
Also, the angular part of the wavefunction will behave closely to $\cos{(3m \theta)}$, in agreement with the results obtained in a previous work where we showed that the angular momentum becomes quantized by $3m$, as also suggested by figures \ref{fig: celda1} and \ref{fig: celda2}. 
In Fig. \ref{fig:potentialA} (b) we present $\bm{A}^2(\bm{r})$. We observe that there are no relevant features that give any indication of a possible confinement. However, such confinement arises when we consider the angular momentum. This is best seen by working near the origin and using polar coordinates. The first equation in (\ref{eq:zero_mode_h2_symmetric_poisson2}) now looks as, 
\begin{equation}
    -\left(\frac{\partial^{2} \psi_{\pm}}{\partial r^{2}}+ \frac{1}{r}\frac{\partial \psi_{\pm}}{\partial r }+\frac{1}{r^{2}}\frac{\partial^{2} \psi_{\pm}}{\partial\theta^{2} }\right) +\alpha^{2}\bm{A}^2(\bm{r}) \psi_{\pm}=0
\end{equation}
As the third term in the Laplacian is the angular momentum, we see that an effective potential appears which contains the moiré symmetric potential part plus the centrifugal barrier, which is a result of the orbital motion of the electron. Elsewhere it was shown \cite{Naumis2023} that the magic angle is given by $\alpha_m \approx 3m/2$ and asymptotically, $L_z \psi_{\pm} \approx m \psi_{\pm}$. Also, we can discard the second term of the Laplacian, as derivatives scale with $\alpha$ inside the boundary layer of the equation \cite{Naumis2023}. We obtain that,
\begin{equation}\label{eq:effHarmOsc}
    -\frac{\partial^{2} \psi_{\pm}}{\partial r^{2}}+\frac{9}{4}m^{2}\left(\frac{1}{r^{2}} +\bm{A}^2(\bm{r})\right) \psi_{\pm} \approx 0
\end{equation}
 A bound state will appear if the effective potential has a minimum. As we also have the condition on the angular part that confines electrons in certain directions, here we will discuss the minimum that results in the $y$ direction. This is seen in Fig. \ref{fig:potencial_eff} where we plot the potentials $\bm{A}^2(0,y), 1/y^{2}$ and the effective one $V_{eff}=1/y^{2}+\bm{A}^2(0,y)$. As seen in the plot, the minima are close to the numerically found limiting confinement centers for the wave functions, indicated in Fig. \ref{fig:potencial_eff} by vertical lines. The minimum can be found from,
\begin{equation}\label{eq:v_eff}
   \left( \frac{dV_{eff}}{dy} \right)_{y=R}=-\frac{2}{R^{3}}+3\sin{(3R/2)}=0
\end{equation}
We found numerically that the minimum is approximately $R\approx 0.88$.
Notice that the obtained minimum is shifted with respect to the numerical obtained value, i.e., the error is $\Delta R \approx 1.047-0.88\approx 0.16$ which is around $15\%$. The reason is that we made several strong approximations like neglecting overlaps between localization centers, the correct shape of the angular part which introduces a factor in the angular momentum, etc. Around the localization center, the effective potential can be approximated with a parabola. Therefore, we obtain an effective harmonic oscillator
equation,
\begin{equation}\label{eq:Harmonic}
    -\frac{\partial^{2} \psi_{\pm}}{\partial y^{2}}+\left(\frac{3m}{2}\right)^{2}\left(V_{eff}(R)+\frac{\omega^{2}(R)}{2}(y-R)^{2}\right)\psi_{\pm} \approx 0
\end{equation}
where the frequency is,
\begin{equation}
\omega^{2}(R)=\left(\frac{d^{2}V_{eff}(y)}{dy^{2}}\right)_{y=R}
=\frac{6}{R^{4}}+\frac{9}{2}\cos{(3R/2)}
\end{equation}
On the other hand, the result from the scaling argument $\sigma$ has an associated frequency $\omega=3\alpha$ (See Ref. \cite{Naumis2023}), as the energy re-scales as $1/\alpha^{2}$. Thus, the scaled frequency is $\omega^{\prime}=\frac{\omega}{\alpha}=3$ and so $\omega^{2}=9$ where primes are omitted.  Therefore, comparing $\omega^{2}=9$ with $\omega^{2}(R)$ at $R=1.047$ we found that $\omega^{2}(R)\approx 9.489$, hence, the error is $\Delta \omega =\omega^{2}-\omega^{2}(R)\approx 0.489$ which is around $5\%$. For $R\approx 0.88$, the frequency is $\omega^{2}(R)\approx 11.121$. The error is $\Delta \omega =\omega^{2}-\omega^{2}(R)\approx 2.121$ which is around $19\%$.

The zero mode can thus be interpreted as the ground state of this effective harmonic oscillator with an energy shift determined by $m^{2}V_{eff}(R)$ and guiding center $R$.  Thus, this explains the Gaussians shapes for the electronic density discussed in the previous section. Finally, it is important to remark that our analysis was made for the $\Gamma$ point. The reason is that such mode is at the top of the band and thus signals the magic angles whenever its corresponding energy goes to zero  \cite{Naumis2022}. At other $\bm{k}$ points, numerical calculations indicate that the wavefunctions also converge towards the same localization center \cite{Naumis2022}. 
This can be easily explained by examining equation (\ref{eq:spinor2}). In the limit $\alpha \rightarrow \infty$, the peaks in reciprocal space satisfy $|\bm{K}_{l,n}| = |l\bm{b}_1 + n\bm{b}_2| \gg |\bm{k}|$ when $l$ and $n$ are much bigger than 1. Consequently, $\bm{k}$ can be safely neglected in all expressions, leading to the collapse of all $\bm{k}$ values into the same equation.

\section{Relationship with the non-Abelian magnetic quantum Hall effect}\label{secNonAbelian}

\begin{figure}[h!]
\includegraphics[scale=.5]{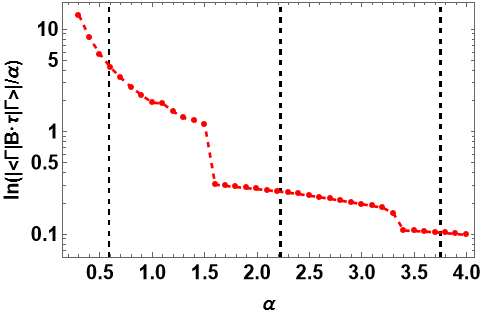}\caption{ Zeeman energy $\log{|\langle \Gamma| \bm{B} \cdot\hat{\bm{\tau}} |\Gamma \rangle/\alpha|}$ as function of $\alpha$ for the zero mode wavefunction at the $\Gamma$-point. As $\alpha$ increases, the Zeeman energy is quite small, and for higher magic angles $\alpha_8$ or $\alpha_9$ can be negligible. Dashed vertical lines indicate the first three magic angles.}
\label{fig:Zeeman}
\end{figure}

In this section, we will explore some interesting connections with non-Abelian magnetic fields. We now write the squared Hamiltonian, 
\begin{equation}
\begin{split}
H^2 &=(-\laplacian+\bm{A}^{2})\tau_0+i\alpha^{2}[A_{x}, A_{y}]\tau_z-2i\alpha\hat{\bm{A}}\cdot\gradient \\
&+\alpha(\partial_{x}\hat{A}_y-\partial_{y}\hat{A}_x)
    \label{eq:H22}
\end{split}
\end{equation}
where $\hat{\tau}_{j}$ (with $j=1,2,3$) is the set of Pauli matrices in the pseudo-spin layer degree, and the identity $2 \time 2$ matrix $\hat{\tau}_{0}$. Moreover, $A_x$ and $A_y$, and its matrices SU(2) versions $\hat{A}_x$ and $\hat{A}_y$ are defined in Appendix A. 
%Below we write in matrix form the diagonal just for illustration purposes, 
%\begin{equation}
%\begin{pmatrix}
%|U(-\bm{r})|^2 & 0\\
%0 & |U(\bm{r})|^2 
%%\end{pmatrix}=\bm{A}^{2}(\bm{r})\tau_0-i[A_x,A_y]\tau_z
%\label{eq:U2_matrix}
%\end{equation}
Written in such way, we can identify the Zeeman coupling energy as,
\begin{equation}
\begin{split}
\hat{F}_{xy} &= \partial_{x}\hat{A}_y-\partial_{y}\hat{A}_x +i\alpha[\hat{A}_x,\hat{A}_y]\\
&= -\hat{\bm{B}}\cdot\hat{\bm{\tau}}+i\alpha[\hat{A}_x,\hat{A}_y]
\label{eq:Zeeman_coupling}
\end{split}
\end{equation}
where upper hats represent matrices. For convenience, we re-scale the spatial coordinates as $\bm{r}^{\prime}=\bm{r}/\alpha$ from where $\gradient^{\prime}=(\alpha\gradient)$ and $(\gradient^{\prime})^{2}=(\alpha\gradient)^2$. The re-scaled position Hamiltonian is, 

\begin{equation}
\begin{split}
(H/\alpha)^2&=(-\laplacian+\bm{A}^{2}(\bm{r}/\alpha))\tau_0+i[A_{x}(\bm{r}/\alpha), A_{y}(\bm{r}/\alpha)]\tau_z \\
& -2i\hat{\bm{A}}(\bm{r}/\alpha)\cdot\gradient-\frac{1}{\alpha}\hat{\bm{B}}(\bm{r}/\alpha)\cdot\hat{\bm{\tau}}
    \label{eq:H2_scaled}
\end{split}
\end{equation}
where now the primes are dropped. As explained in Appendix A, the strong confinement of electrons allows to suppose an almost uniform magnetic field. This is as also seen in the effective eq. (\ref{eq:effHarmOsc}). Therefore, we can write $\bm{A}\cdot\hat{\bm{p}}\approx -\bm{B}\cdot\hat{\bm{L}}$ where $\hat{\bm{L}}$ is the total angular moment. Under such simplification, the re-scaled Hamiltonian is, 

\begin{equation}
\begin{split}
\hat{H}^2&=\overbrace{(-\laplacian+\bm{A}^{2}(\bm{r}/\alpha))\tau_0}^\text{diagonal energy}+\underbrace{i[A_{x}(\bm{r}/\alpha), A_{y}(\bm{r}/\alpha)]\tau_z}_\text{non-Abelian energy}\\
&\underbrace{-\hat{\bm{B}}(\bm{r}/\alpha)\cdot(2\hat{\bm{L}}+\frac{\bm{e}_z}{\alpha})}_\text{off-diagonal energy}
\label{eq:H2_Magnetic_off_diagonal2}
\end{split}
\end{equation}
Note that only the last term depends on $\alpha$ and taking the asymptotic limit $\alpha\rightarrow\infty$ we have that the Zeeman energy $-\frac{1}{\alpha}\bm{B}(\bm{r}/\alpha)\cdot\bm{\tau}\rightarrow 0$. This fact is corroborated in Fig. \ref{fig:Zeeman}, where it can be observed that for the first magic angle, the expected value of the Zeeman energy scaled by $\alpha$ is significant. However, for the third magic angle, it is very small, around $0.1$ on the logarithmic scale. Therefore, it is expected to be similarly small for higher magic angles, and neglecting it should not significantly impact the results. 
Thus, in the asymptotic limit $\alpha\rightarrow\infty$, $2\hat{\bm{B}}(\bm{r}/\alpha)\cdot\hat{\bm{L}}>>\hat{\bm{B}}(\bm{r}/\alpha)\cdot\bm{e}_z/\alpha$, i.e., $E_{Magnetic}>>E_{Zeeman}$.  Hence, the Hamiltonian in this limit can be simplified into,
\begin{equation}
\hat{H}^2=\overbrace{(\bm{p}+\hat{\bm{A}}(\bm{r}/\alpha))^{2}}^\text{$C_3$ magnetic field}+\underbrace{i[A_{x}(\bm{r}/\alpha), A_{y}(\bm{r}/\alpha)]\tau_z}_\text{non-Abelian operator}
    \label{eq:H2_scaled_limit}
\end{equation}
where $\hat{H}^2=(H/\alpha)^2$ and $\bm{p}=-i\gradient$ is the canonical momentum operator. Accordingly, $\hat{H}^2$ it's expected to have a non-Abelian QHE.\\

Let us know discuss how the magic angle order parity enters inside the orbital magnetic energy related to the angular momentum chirality.  To understand this we start by writing the zero mode equation $H^2\psi(\bm{r})=0$ together with eq. (\ref{eq:H2_Magnetic_off_diagonal2}) at the $\Gamma$-point, where $\psi_2(\bm{r})=i\mu_{\alpha}\psi_1(-\bm{r})$. Using the results of Appendix A in the limit $\alpha\rightarrow\infty$, such that the wave function at the $\Gamma$-point is strongly confined, we obtain,

\begin{equation}
\begin{split}
(-\laplacian &+ \bm{A}^2(\bm{r}/\alpha) + \Delta(\bm{r}/\alpha))\psi_{1}(\bm{r})\\
&-2i\mu_{\alpha}\bm{B}(\bm{r}/\alpha)\cdot\hat{\bm{L}}\psi_{1}(\bm{-r}) =0 
    \label{eq:equation1_zeromode}
\end{split}
\end{equation}
 The corresponding expected values over the zero mode wavefunction at the $\Gamma$-point are, 

\begin{equation}
\begin{split}
\langle\Gamma| T(\bm{r}/\alpha)|\Gamma  \rangle + \langle\Gamma| \bm{A}^2(\bm{r}/\alpha) |\Gamma\rangle -2i\mu_{\alpha}\langle\Gamma| \bm{B}(\bm{r}/\alpha)\cdot\hat{\bm{L}} |\Gamma\rangle
=0 
    \label{eq:equation2_zeromode}
\end{split}
\end{equation}
where $T(\bm{r}/\alpha)$ is the kinetic energy, i.e., minus the Laplacian, and we have used that the anti-symmetric potential is canceled inside the unit cell $\langle\Gamma| \Delta(\bm{r}/\alpha)  |\Gamma\rangle=0$ (see Fig. \ref{fig:potentialA}(a)).  At magic angles we can use the energy equipartition  
 found in a previous work \cite{Naumis2022}, from where $\langle\Gamma| T(\bm{r}/\alpha)  |\Gamma\rangle=\langle\Gamma| \bm{A}^2  (\bm{r}/\alpha)|\Gamma\rangle$. Thus, 
\begin{equation}
\begin{split}
\langle\Gamma| \bm{A}^2(\bm{r}/\alpha) |\Gamma\rangle -i\mu_{\alpha}\langle\Gamma| \bm{B}(\bm{r}/\alpha)\cdot\hat{\bm{L}} |\Gamma\rangle =0 
\label{eq:equation3_zeromode}
\end{split}
\end{equation}
where is important to note that,
\begin{equation}
\begin{split}
-i\mu_{\alpha}\bm{B}(\bm{r}/\alpha)\cdot\hat{\bm{L}}&=-i\sum_{\nu}(-i)e^{-i\bm{q}_{\nu}\cdot\bm{r}/\alpha}\bm{e}_z\cdot(\mu_{\alpha}\bm{q}_{\nu}\times\hat{\bm{p}})\\
&=-\sum_{\nu}e^{-i\bm{q}_{\nu}\cdot\bm{r}/\alpha}\bm{e}_z\cdot(\mu_{\alpha}\bm{\hat{L}}_{\nu})\\
&=-\sum_{\nu}\bm{B}_{\nu}(\bm{r}/\alpha)
\cdot(\mu_{\alpha}\hat{\bm{L}}_{\nu})
\label{eq:magnetic_relation}
\end{split}
\end{equation}
where $\bm{B}_{\nu}(\pm\bm{r}/\alpha)=\pm ie^{\pm i\bm{q}_{\nu}\cdot\bm{r}/\alpha}$ and we defined, 
\begin{equation}
\begin{split}
\hat{\bm{M}}_{\nu}=\mu_{\alpha}\bm{\hat{L}}_{\nu}
\label{eq:magnetic_momentum}
\end{split}
\end{equation}
as the pseudo-magnetic orbital momentum at the direction $\nu$, with $\bm{\hat{L}}_{\nu}=\bm{q}_{\nu}\times\hat{\bm{p}}$ a kind of angular momentum operator. We can understand its origin as a consequence of the strong confinement as in the angular momentum $\bm{\hat{L}}_{z}=\bm{r}\times\bm{p}$, $\bm{r}$ takes only values different from zero at $\bm{r} \approx \bm{q}_{\nu}$. Therefore, we can interpret $\hat{\bm{L}}_{\nu}$ as the contribution to the angular momentum of each confinement center, as these centers are not in the origin of coordinates. Such observation was empirically made by analyzing the numerical data in a previous paper \cite{Naumis2023}.  In the asymptotic limit $\alpha\rightarrow\infty$ we have that \cite{Naumis2022} $\langle \Gamma|\bm{A}^2(\bm{r}/\alpha) |\Gamma\rangle \rightarrow 1$ from where, 

\begin{equation}
\begin{split}
1 -\int d^{2}\bm{r}\psi^{\dagger}_{1}(\bm{r})\sum_{\nu}e^{-i\bm{q}_{\nu}\cdot\bm{r}/\alpha}\bm{e}_z\cdot(\mu_{\alpha}\bm{\hat{L}}_{\nu})\psi_{1}(-\bm{r})=0 
\label{eq:equation4_zeromode}
\end{split}
\end{equation}

therefore, 
\begin{equation}
 \begin{split}
1-&\mu_{\alpha}\bm{e}_z\cdot\sum_{\nu}\int d^{2}\bm{r}\psi^{\dagger}_{1}(\bm{r})e^{-i\bm{q}_{\nu}\cdot\bm{r}/\alpha}\bm{\hat{L}}_{\nu}\psi_{1}(-\bm{r})\\
 &=1-\mu_{\alpha}|\bm{e}_z|^{2}\sum_{\nu}(\frac{\mu_{\alpha}}{3})\\
 &=1-\mu^2_{\alpha}=0
  \label{eq:angular_zeromode_mu}
 \end{split}
\end{equation}
where are used natural units $e=\hbar=1$ and rescaled energies $1/\alpha^2$, normalized over the moiré unit cell area. Each contribution of plane waves in the sum contributes $1/3$ to the integral, i.e., 
\begin{equation}
 \begin{split}
\frac{1}{\alpha A_M}\langle \psi_{1}(\bm{r})| \bm{B}_{\nu}(\bm{r}/\alpha) \cdot \hat{\bm{L}}_{\nu} |\psi_{1}(-\bm{r})\rangle=\frac{\mu_{\alpha}}{3}
\label{eq:angular_zeromode_mu_one_component}
\end{split}
\end{equation}
\\
where $A_M=8\pi^2/(3\sqrt{3})$ is the normalized moiré unit cell area. This proves 
that parity and the three directional components of the angular momentum are essential
to satisfy the magic angle condition. Moreover. eq. (\ref{eq:magnetic_momentum}) indicates that the parity is related with the chirality of the magnetic energy.

\begin{figure}[h!]
\includegraphics[scale=.50]{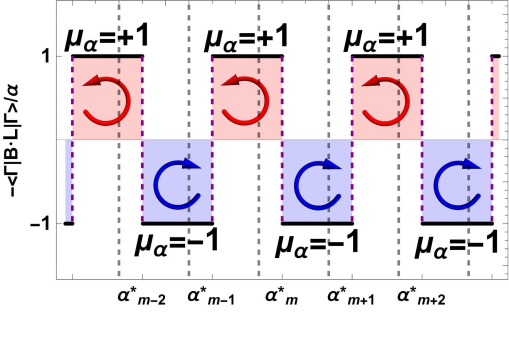}\caption{
Orbital magnetic energy $-\langle\Gamma| \bm{B} \cdot \hat{\bm{L}} |\Gamma\rangle/\alpha$ as function of $\alpha$ in the limit $\alpha\rightarrow\infty$ for the zero mode wavefunction at $\Gamma$-point, obtained from the numerical data of the wave function as in previous works \cite{Naumis2022,Naumis2023}.  Vertical dashed lines (black) indicate magic angles. The red and blue arrows indicate the magnetic orbital rotation, $\mu_{\alpha}=+1$ is counter-clockwise and $\mu_{\alpha}=-1$ is clockwise rotation. Here are considered scaled coordinates $\bm{r}^{\prime}=\bm{r}/\alpha$, when $\alpha\rightarrow\infty$ approximately $\alpha\approx 3m$ where $m>>1$ is the order of the magic angle and $-\langle\Gamma| \bm{B}(\bm{r}/\alpha) \cdot \hat{\bm{L}} |\Gamma\rangle/\alpha\approx\mu_{\alpha}$. The transition points $\alpha^{*}_{m}$, in between magic angles $\alpha_m$ and $\alpha_{m+1}$, occurs when the flat band touches the upper band generating a transition and consequently changes the magnetic orbital orientation. These touching points relate to the magic angle recurrence. Similarly, in the other layer $\bm{B}(\bm{r}/\alpha)\rightarrow \bm{B}(-\bm{r}/\alpha)$. }
\label{fig:magnetic_parity}
\end{figure}

To corroborate the chirality of the magnetic energy, in Fig. \ref{fig:magnetic_parity}, we plot $\langle\Gamma|\bm{B}(\bm{r}/\alpha)\cdot \hat{\bm{L}} |\Gamma\rangle/\alpha$ versus $\alpha$ at the $\Gamma$-point as obtained from the numerical data of the wave function,  by using techniques described in previous works \cite{Naumis2022,Naumis2023}. 
In the $y$-axis, this magnetic energy jumps from $\mu_{\alpha}=+1\rightarrow -1$ or vice-versa. Because we rescaled the coordinates, the energy is also rescaled as $E^{\prime 2 }=(E/\alpha)^2$, and thus the result does not depend on $\alpha$.

Fig. \ref{fig:magnetic_parity} also shows the relation between $\mu_{\alpha}=+1$ counter-clockwise rotation (red arrows) and $\mu_{\alpha}=-1$ clockwise rotation (blue arrows) as the $z$-component rotation of the magnetic angular momentum. The values $\alpha^{*}_m$ indicate the intermediate values between magic angles $\alpha_m$ and $\alpha_{m+1}$. At these special values, the gap closes and the zero mode hybridizes with its neighbor upper band changing the chirality of the angular momentum. 

Thus, an important characteristic of TBG  is the gap closing in between magic angles due to the hybridization of the lowest band with its neighbor upper band. This is a crucial condition because is a transition that changes the chirality of the angular momentum and the magic angle order parity $\mu_{\alpha}=\pm 1$. At the same time, on each gap closing appears a new quanta of  angular momentum, and consequently, the magnetic angular momentum increases as $\alpha\rightarrow\infty$.

So far, in this analysis is clear that parity of the wavefunction and the sign $\mu_{\alpha}$ plays a crucial role in the energetic balance for magic angles flat bands, nevertheless, only at higher magic angles does the wave function reaches a purely symmetric or anti-symmetric solution and in this way, the angular momentum quantum number and the magic angle order parity governs the physics behind flat bands.

\section{Competition of Non-Abelian and Abelian fields}\label{secBeta}

The chiral TBG model is quite interesting and exhibits remarkable properties due to its non-Abelian nature introduced by the coupling potential $U(\bm{r})$ between layers \cite{Guinea2012,Naumis2023r}. In fact, flat bands and superconductivity in TBG are consequences of the underlying pseudo-magnetic fields generated by the twist angle. However, what if we could tune non-Abelian fields to become Abelian using an artificial parameter? How would this modification affect the periodicity and quantization of magic angles? To explore this effect, we can define a new coupling potential as follows,

\begin{equation}
\begin{split}
U_{\beta}(\bm{r})=U(\bm{r})+\beta U(-\bm{r})
\label{eq:U_beta}
\end{split}
\end{equation}
where $\beta$ is the artificial parameter that controls the non-Abelian nature of TBG. Suppose that $\beta\in [0,1]$, with $\beta=0$ we recovered the cTBG case while $\beta=1$ is presumably an Abelian case. Using this new potential we can write a new Hamiltonian as, 
\begin{equation}
\begin{split}
\mathcal{H}_{\beta}
&=\begin{pmatrix} 
0 & D^{\ast}_{\beta}(-\bm{r})\\
 D_{\beta}(\bm{r}) & 0
  \end{pmatrix}  \\
\end{split} 
\label{H_initial_beta}
\end{equation}
where the zero mode operator is, 
\begin{equation}
\begin{split}
D_{\beta}(\bm{r})&=\begin{pmatrix} 
-i\Bar{\partial} & \alpha U_{\beta}(\bm{r})\\
  \alpha U_{\beta}(-\bm{r}) & -i\Bar{\partial} 
  \end{pmatrix}  \\
\end{split} 
\end{equation}
\\

The Abelian case $\beta=1$ gives, 
\begin{equation}
\begin{split}
D_{1}(\bm{r})&=\begin{pmatrix} 
-i\Bar{\partial} & 0\\
  0 & -i\Bar{\partial} 
  \end{pmatrix}+\begin{pmatrix} 
0 & \alpha U_{1}(\bm{r})\\
  \alpha U_{1}(-\bm{r}) & 0 
  \end{pmatrix}  \\
\end{split} 
\end{equation}
however, $U_{1}(-\bm{r})=U_{1}(\bm{r})$ so, 
\begin{equation}
\begin{split}
D_{1}(\bm{r})&=-i\Bar{\partial}\hat{\tau}_{0}+\alpha U_{1}(\bm{r})\hat{\tau}_{x} \\
\end{split} 
\end{equation}
where $U_{1}(\bm{r})=2\sum_{\nu}e^{i(\nu-1)\phi}\cos{(\bm{q}_{\nu}\cdot\bm{r})}$ is the symmetric coupling potential. Now is clear from these expressions that the vector potential commute and the initial $SU(2)$ gauge field change to a $U(1)$ field. \\

\begin{figure}[h!]
\includegraphics[scale=.65]{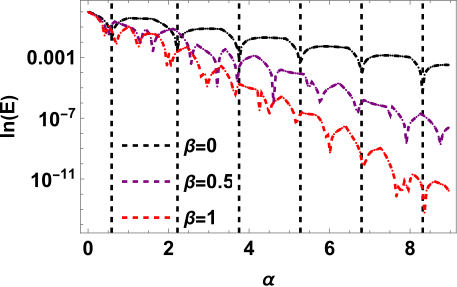}\caption{Energy $E$, in log scale, as a function of $\alpha$ at the $\bm{\Gamma}$-point. The $\beta$ parameter transforms the original chiral model with a non-Abelian nature to an Abelian system. In the curve $\beta=1$, the off-diagonal term is proportional to $\hat{\tau}_x$ and there is no well-defined $3/2$ magic angle recurrence as for the cTBG ($\beta=0$). Vertical lines indicate magic angles.}
\label{fig:beta_energy}
\end{figure}

Fig. \ref{fig:beta_energy} shows the zero energy mode in log scale as a function of $\alpha$ for different values of $\beta$. 
The non-Abelian structure of cTBG clearly plays a vital role in magic angle recurrence. Interestingly, even at $\beta=1$  it exhibits a decaying behavior; however, it does not have a well-defined $3/2$ magic angle recurrence rule. Furthermore, when $\beta=0\rightarrow1$ the band gap has an extra squeezing as $\Delta\sim\Delta_{\alpha}e^{-C\beta}$ where $C$ is a scaling constant and $\Delta_{\alpha}$ is the original band gap of cTBG independent of the parameter $\beta$. 

\section{Conclusion}\label{secConclusion}

In this work, we studied twisted bilayer graphene (TBG) at small magic angles to understand the properties of the electron wave functions. We corroborated that zero mode states converge into coherent Landau states with minimal dispersion. In reciprocal space, they have the same shape (almost Gaussian)  as in real space but with inverted parameters. These coherent states exhibit minimal dispersion with a standard deviation in reciprocal space of $\sigma_k=\sqrt{3\alpha/2\pi}$ as $\alpha$ approaches infinity.

Importantly, as $\alpha$ approaches infinity, the zero mode equation decouples into its symmetric and anti-symmetric components. Exploiting this property and the squared Hamiltonian, we have elucidated the reason for the confinement of the electronic wavefunction as $\alpha$ tends to infinity. Specifically, this confinement arises from the interplay between the squared norm of the moiré potential and the quantized orbital motion of electrons, resulting in the formation of a quantum well. Inside this well, an effective harmonic oscillator is identified, giving rise to Landau levels. 

As the squared Hamiltonian gives rise to an effective quantum oscillator, we also showed how to relate it with the non-Abelian quantum Hall effect. Then we defined a magnetic and Zeeman energy. The Zeeman energy is negible for high order magic angles, while the magnetic term can be interpreted as an orbital magnetic energy with a well defined chirality. This highlight the important role of the $\Gamma$ point wave function parity, as it changes at each gap closing. Finally, we also altered the non-Abelian intrinsic behavior of TBG to see how the $3/2$ quantization rule of flat bands is destroyed by such artifact.

 Therefore, we conclude that the relationship with between TBG physics and  the QHE is not coincidental.  Our recent analytical work on flat bands in graphene without twists has also confirmed such conclusion in a very clear and concise way \cite{Elias2023}.

%Many works in the field of unconventional superconductivity have observed similar behavior, where a change in chirality in the magnetic order and the breaking of parity symmetry play crucial roles \cite{Yanase2018,Kharzeev2020,Macdonald2021, Youichi2022}. It might be interesting to explore analogies with TBG in these directions. \\

This work was supported by  (L.A.N.L. and G.G.N.) and CONAHCyT project 1564464. Leonardo Navarro is supported by a CONAHCyT PhD schoolarship. We thank  Eslam Khalaf at Harvard University (now at Texas University) for valuable comments on the section concerning the artificial potential.

\section{Appendix A: Non-Abelian pseudo-magnetic field and angular momentum}\label{appendixA}
As explained before, electrons in TBG behaves like a  $SU(2)$ non-Abelian pseudo-magnetic vector potential. In matrix notation, it follows that, 
\begin{equation}
\begin{split}
\hat{\bm{A}}=(\hat{A}_x,\hat{A}_y)
\label{eq:A_vector}
\end{split}
\end{equation}
with $\hat{A}_x=A_{1, x}\hat{\tau}_{1}+A_{2, x}\hat{\tau}_{2}$ and $\hat{A}_y=A_{1, y}\hat{\tau}_{1}+A_{2, y}\hat{\tau}_{2}$ where we used the set of Pauli matrices $\hat{\tau}_{j}$ (with $j=1,2,3$) in the pseudo-spin layer degree, and the identity matrix $\hat{\tau}_{0}$. Explicitly, the components of $\hat{\bm{A}}$ are,
\begin{equation}
\begin{split}
A_{1, x}=\sum_{\nu}\cos{(\bm{q}_{\nu}\cdot\bm{r})}\bm{q}^{\perp,x}_{\nu},\\
A_{2, x}=\sum_{\nu}\cos{(\bm{q}_{\nu}\cdot\bm{r})}\bm{q}^{\perp,y}_{\nu},\\
A_{1, y}=\sum_{\nu}\sin{(\bm{q}_{\nu}\cdot\bm{r})}\bm{q}^{\perp,x}_{\nu},\\
A_{2, y}=\sum_{\nu}\sin{(\bm{q}_{\nu}\cdot\bm{r})}\bm{q}^{\perp,y}_{\nu}.
\label{eq:A_vector_components}
\end{split}
\end{equation}
Note that $\hat{\bm{A}}$ is non-Abelian as follows from the fact that $[\hat{\bm{A}}_{\nu},\hat{\bm{A}}_{\eta}]\neq 0$ for $\nu\neq\eta$.  On the other hand, the off-diagonal terms of $H^{2}$ related to the angular momentum and interlayer currents \cite {Naumis2022} have two contributions,

\begin{equation}
\begin{split}
\nabla\times\bm{A}_{\pm}=\bm{B}_{\pm}
\label{eq:off_diagonal2}
\end{split}
\end{equation}
where $\bm{B}_{\pm}$ represents a pseudo-magnetic field while the other term is, 

\begin{equation}
\begin{split}
-2i\bm{A}_{\pm}\cdot\nabla &= -2\bm{B}_{\pm}\cdot\hat{\bm{L}}\\
\label{eq:off_diagonal1}
\end{split}
\end{equation}
Explicitly, we have that, 
\begin{equation}
\bm{A}(\pm\bm{r})\cdot\hat{\bm{p}}=-\sum_{\nu}B_{\nu}(\pm\bm{r})\bm{e}_z\cdot(\bm{q}_{\nu}\times\hat{\bm{p}})
    \label{eq:A_pot_vector}
\end{equation}
where is convenient to define $\bm{q}_{\nu}\times\hat{\bm{p}}= \hat{\bm{L}}_{\nu}$
as an operator similar to the angular momentum at the direction $\nu$, defined by the reciprocal vectors $\bm{q}_{\nu}$. We can interpret $\hat{\bm{L}}_{\nu}$ as the contribution to the angular momentum of each confinement center as $\bm{r} \approx \bm{q}_{\nu}$.  Accordingly, we can re-express the last relation in a compact form as, 
\begin{equation}
\bm{A}(\pm\bm{r})\cdot\hat{\bm{p}}=-\sum_{\nu}\bm{B}_{\nu}(\pm\bm{r})\cdot\hat{\bm{L}}_{\nu}
    \label{eq:A_pot_vector_L}
\end{equation}
where $\bm{A}(\pm\bm{r})=\sum_{\nu}e^{\pm i \bm{q}_{\nu}\cdot\bm{r}}\bm{q}^{\perp}_{\nu}$ with $\bm{q}^{\perp}_{\nu}=\bm{q}_{\nu}\times\bm{e}_z$.  The well known relation $\bm{A}\cdot\hat{\bm{p}}=-\bm{B}\cdot\hat{\bm{L}}$ 
is used here and comes from an uniform and symmetric gauge magnetic vector potential which can be expressed as $\bm{A}=-\frac{1}{2}\bm{r}\times\bm{B}$, where $\bm{r}$ is the position vector and $\bm{B}$ is the magnetic field. It can be used due to the confinement nature of the wave function which allows to suppose a local uniform magnetic field in the spirit of eq. (\ref{eq:Harmonic}). 

Clearly we need to recognize the differences in cTBG compared to the conventional QHE in a radial symmetric potential, i.e., cTBG has  a $C_3$ symmetry and the periodicity of the superlattice. Moreover, the pseudo-magnetic fields are position-dependent, and therefore, spatially inhomogeneous. Surprisingly, despite these differences, cTBG satisfies this magnetic property due to the local Abelian features induced by confinement.  

Hence, Eq. (\ref{eq:A_pot_vector_L}) is analogous to the relation $\bm{A}\cdot\hat{\bm{p}}=-\bm{B}\cdot\hat{\bm{L}}$ used in symmetric gauge magnetic fields. Note in eq. (\ref{eq:A_pot_vector_L}) that the direct product between the pseudo-magnetic field and the angular momentum is a superposition of three-plane waves. 
This off-diagonal operator is quite important for engineering flat bands at magic angles, moreover, introduces the magic angle order parity in the energy equipartition rule balance for flat bands.
\\

On the other hand, the squared TBG system is a $2\times 2$ matrix operator where the layer degree of freedom introduces $SU(2)$ Pauli matrices $\bm{\tau}$, in this manner, is convenient to re-express the off-diagonal operator using matrices to consider the effect of both layers, from where it follows that, 

\begin{equation}
-2i\hat{\bm{A}}\cdot\gradient=
\begin{pmatrix}
0 & 2\sum_{\nu}e^{-i\bm{q}_{\nu}\cdot\bm{r}}\bm{q}^{\perp}_{\nu}\cdot\hat{\bm{p}}\\
2\sum_{\nu}e^{i\bm{q}_{\nu}\cdot\bm{r}}\bm{q}^{\perp}_{\nu}\cdot\hat{\bm{p}} & 0 
\end{pmatrix}
\label{eq:A_matrices}
\end{equation}
since $\hat{\bm{A}}\cdot\hat{\bm{p}} \approx -\hat{\bm{B}}\cdot\hat{\bm{L}}$ follows that, 

\begin{equation}
\begin{split}
-2i\hat{\bm{A}}\cdot\gradient &=
\begin{pmatrix}
0 & 2 A(\bm{r})\cdot\hat{\bm{p}}\\
2 A(-\bm{r})\cdot\hat{\bm{p}} & 0 
\end{pmatrix}\\
&=2\begin{pmatrix}
0 & -B(\bm{r})\cdot\hat{\bm{p}}\\
-B(-\bm{r})\cdot\hat{\bm{p}} & 0 
\end{pmatrix}
\label{eq:A_matrices2}
\end{split}
\end{equation}
This operator is responsible for coupling the layers with pseudo-magnetic potentials $B(\bm{r})$ (layer 1) and $B(-\bm{r})$ (layer 2). This matrix form gives us more insight into the non-Abelian nature of the pseudo-magnetic potentials related to the $SU(2)$ layer degree of freedom. 
\\

\section{Appendix B: Symmetrized zero mode equation at the asymptotic limit $\alpha\rightarrow\infty$}\label{appendixB}

As was mentioned in sec. \ref{secConfinement}, at the asymptotic limit the zero mode equation is decoupled into two separate equations as follows, 
\begin{equation}
(-\laplacian+\alpha^{2}\bm{A}^2-i\mu_{\alpha}\alpha( -2i\mathcal{A}_{\mp}\cdot\nabla+\nabla\times\mathcal{A}_{\mp}))\psi_{\pm} \approx 0
\label{eq:zero_mode_h2_symmetric_decoupled_Appendix}
\end{equation}
\begin{equation}
(\alpha^{2}\Delta-i\mu_{\alpha}\alpha( -2i\mathcal{A}_{\pm}\cdot\nabla+\nabla\times\mathcal{A}_{\pm}))\psi_{\mp} \approx 0
\label{eq:zero_mode_h2_symmetric_decoupled2_Appendix}
\end{equation}
From where if we consider scaling of the spatial coordinates as, $\bm{r}^{\prime}=\bm{r}/\alpha$ and therefore, $\gradient^{\prime}=(\alpha\gradient)$ and $(\gradient^{\prime})^{2}=(\alpha\gradient)^2$ it follows that energy scale proportional to $\alpha^{2}$, thus eq. (\ref{eq:zero_mode_h2_symmetric_decoupled_Appendix}) and eq. (\ref{eq:zero_mode_h2_symmetric_decoupled2_Appendix}) changes as, 
\begin{equation}
(-\laplacian+\bm{A}^2(\bm{r}/\alpha)-2\mu_{\alpha}\mathcal{A}_{\mp}(\bm{r}/\alpha)\cdot\nabla)\psi_{\pm} \approx 0
\label{eq:zero_mode_h2_symmetric_decoupled_Appendix_alpha}
\end{equation}
\begin{equation}
(\Delta(\bm{r}/\alpha)-2\mu_{\alpha}\mathcal{A}_{\pm}(\bm{r}/\alpha)\cdot\nabla)\psi_{\mp} \approx 0
\label{eq:zero_mode_h2_symmetric_decoupled2_Appendix_alpha}
\end{equation}
where the term $\nabla\times\mathcal{A}_{\pm}(\bm{r}/\alpha)=\frac{1}{\alpha}\mathcal{B}_{\pm}\rightarrow0$ as $\alpha\rightarrow\infty$. From eq. (\ref{eq:zero_mode_h2_symmetric_decoupled2_Appendix_alpha}) follows that, 
\begin{equation}
\Delta(\bm{r}/\alpha)\psi_{\mp}=2\mu_{\alpha}\mathcal{A}_{\pm}(\bm{r}/\alpha)\cdot\nabla\psi_{\mp}
\label{eq:zero_mode_h2_symmetric_decoupled2_Appendix_alpha3}
\end{equation}
thus, substituting eq. (\ref{eq:zero_mode_h2_symmetric_decoupled2_Appendix_alpha3}) into eq. (\ref{eq:zero_mode_h2_symmetric_decoupled_Appendix_alpha}) is easy to show that, 
\begin{equation}
(-\laplacian+\bm{A}^2(\bm{r}/\alpha)-\Delta(\bm{r}/\alpha))\psi_{\pm} \approx 0
\label{eq:zero_mode_h2_symmetric_decoupled_Appendix_alpha_laplace}
\end{equation}
From this last expression is clear that we can decouple into two separate equations, 
\begin{equation}
(-\laplacian+\bm{A}^2(\bm{r}/\alpha))\psi_{\pm} \approx 0
\label{eq:zero_mode_h2_symmetric_decoupled_Appendix_alpha_laplace1}
\end{equation}
and
\begin{equation}
\Delta(\bm{r}/\alpha)\psi_{\pm} \approx 0
\label{eq:zero_mode_h2_symmetric_decoupled_Appendix_alpha_laplace2}
\end{equation}
These equations give the localization behavior in the asymptotic limit $ \alpha \rightarrow \infty$. Both eqns. (\ref{eq:zero_mode_h2_symmetric_decoupled_Appendix_alpha_laplace1}) and (\ref{eq:zero_mode_h2_symmetric_decoupled_Appendix_alpha_laplace2}) gives information related to the radial and angular confinement position, respectively. In particular, the angular directions are defined by $\Delta(\bm{r})=0$ giving confinement paths along the unitary vectors $\pm\bm{q}_{\nu}$, this is analogous to saying that $[A_x, A_y]=0$, therefore, the electronic wave function is locally Abelian. In this manner, cTBG can be interpreted at the asymptotic limit $\alpha\rightarrow\infty$ as an effective quasi-1D system along these preferential directions.

%\bibliography{references.bib}
%apsrev4-2.bst 2019-01-14 (MD) hand-edited version of apsrev4-1.bst
%Control: key (0)
%Control: author (72) initials jnrlst
%Control: editor formatted (1) identically to author
%Control: production of article title (-1) disabled
%Control: page (0) single
%Control: year (1) truncated
%Control: production of eprint (0) enabled
%
\end{document}